\documentclass[conference]{IEEEtran}
\IEEEoverridecommandlockouts
\usepackage{cite}
\usepackage{amsmath,amssymb,amsfonts}
\usepackage{algorithmic}
\usepackage{graphicx}
\usepackage{textcomp}
\usepackage{xcolor}
\def\BibTeX{{\rm B\kern-.05em{\sc i\kern-.025em b}\kern-.08em
    T\kern-.1667em\lower.7ex\hbox{E}\kern-.125emX}}
\begin{document}

\title{An Ecosystem Approach to Ethical AI and Data Use: Experimental Reflections\\
\thanks{This research is supported by the National Research Foundation, Singapore under its Emerging Areas Research Projects (EARP) Funding Initiative. Any opinions, findings and conclusions or recommendations expressed in this material are those of the author(s) and do not reflect the views of National Research Foundation, Singapore.}
}

\author{\IEEEauthorblockN{Mark Findlay}
\IEEEauthorblockA{\textit{Centre for AI and Data Governance} \\
\textit{School of Law, Singapore Management University}\\
Singapore \\
markfindlay@smu.edu.sg}
\and
\IEEEauthorblockN{Josephine Seah}
\IEEEauthorblockA{\textit{Centre for AI and Data Governance} \\
\textit{School of Law, Singapore Management University}\\
Singapore \\
shseah@smu.edu.sg}
}

\IEEEoverridecommandlockouts
\IEEEpubid{\makebox[\columnwidth]{
\copyright2020
IEEE \hfill} \hspace{\columnsep}\makebox[\columnwidth]{ }}

\maketitle

\begin{abstract}
While we have witnessed a rapid growth of ethics documents meant to guide artificial intelligence (AI) development, the promotion of AI ethics has nonetheless proceeded with little input from AI practitioners themselves. Given the proliferation of AI for Social Good initiatives, this is an emerging gap that needs to be addressed in order to develop more meaningful ethical approaches to AI use and development. This paper offers a methodology—a ‘shared fairness’ approach—aimed at identifying AI practitioners’ needs when it comes to confronting and resolving ethical challenges and to find a third space where their operational language can be married with that of the more abstract principles that presently remain at the periphery of their work experiences. We offer a grassroots approach to operational ethics based on dialog and mutualised responsibility: this methodology is centred around conversations intended to elicit practitioners perceived ethical attribution and distribution over key value-laden operational decisions, to identify when these decisions arise and what ethical challenges they confront, and to engage in a language of ethics and responsibility which enables practitioners to internalise ethical responsibility. The methodology bridges responsibility imbalances that rest in structural decision-making power and elite technical knowledge, by commencing with personal, facilitated conversations, returning the ethical discourse to those meant to give it meaning at the sharp end of the ecosystem. Our primary contribution is to add to the recent literature seeking to bring AI practitioners' experiences to the fore by offering a methodology for understanding how ethics manifests as a relational and interdependent sociotechnical practice in their work. 

\end{abstract}


\section{Introduction}
\begin{quote}
    But on a contract and a project for another company, the number one thing that is driving choices is meeting terms of the contract, and there is little room for thinking about ethics, we’re not breaking rules, but it boils down to meeting deadlines and getting things ready – rushed – as long as nothing is flagrant you do what needs to get done.\footnote{Comment made by a workshop participant, March 2020.}
\end{quote}

The concept of AI for Social Good (AI4SG) has been adopted widely by developers within the AI community\cite{floridi2020design}; while giant tech companies--themselves strong promoters of the principled ethics approach--highlight the capacity of AI to contribute to the achievement of the United Nations' Sustainable Development Goals (SDGs). Google has, for example, teamed up with the United Nations' Economic and Social Commission for Asia and the Pacific (ESCAP) to sponsor research into AI4SG, with a focus on achieving the SDGs in the Asia-Pacific region. Yet, both AI and big data remain under-regulated phenomena. In response to rising concerns of unethical AI development, the prevailing approach from governments, intergovernmental organisations, and big tech firms has been to roll out a ‘new’ vocabulary of ethics and responsibility\cite{jobin2019global}. This approach relies on broadly-stated ethics frameworks intended to moralise market dynamics and elicit socially responsible behaviour among AI developers and users. The ethical frameworks and codes are intended to engender trust across communities. Nonetheless, these codes have recently come under criticism for deflecting responsibility by generating a smokescreen of agreeable but fuzzy principles\cite{ochigame2019invention,mittelstadt2019principles,hagendorff2020ethics}. 

Despite these limitations, we suggest that ethics--applied and evaluated contextually--continue to be an important framework against which crucial decisions are made. To give ethics sufficient regulatory bite, political, economic, social, operational, and sustainability externalities must be recognised as having an equally significant place as determinants of necessary behaviour. As we show below, if AI practitioners are limited in their actions by organisational power hierarchies and contract pressures, then the intended operational influence of ethics is compromised.  

This paper engages with the ethical regulation of AI at three levels. The first is to generate and share emerging conversations about ethics with AI practitioners so that mutual responsibility in attribution and distribution of ethical considerations in key-decision sites can be evaluated. As our empirical experience reveals, if ethics is being blind-sided by predictable and recurrent operational pressures and compromises, then this information needs feeding into a pragmatic evaluation of the regulatory promises of ‘Ethical AI’. This experience also reveals that the general nature, form and comprehension of principle-structured ethical discourse is often an impediment to spontaneous, engaged and informed conversations around ethics in use case contexts. Second, and embedded within sustainable conversations, is the pressing requirement to interrogate a meaningful language and understanding of ethics across all stages and decision domains of the AI ecosystem, and thereby create the possibility of evaluating ethics as an inclusive regulatory frame in work-life experience. For this purpose, ethics, AI, big data and human agency are seen as a communal enterprise. The responsibility for devising, agreeing and applying a relevant ethical language is mutualised throughout the AI ecosystem and on to its varied applications. Finally, the project aims to prioritise a central element within the ethical panoply, that being fairness, and attempt to model ways in which fairness can not only be shared but effectively and influentially directed to AI and big data applications so that human dignity is maintained and maximised.

The theoretical purpose of our contribution is as such: if ascription to ethical principles prevents social harm arising from the development and use of AI, this needs to be a holistic enterprise to maximise its regulatory influence across the AI ecosystem. Ethics guidelines, if they only have contained or exclusive sectoral impact in the ecosystem will be limited in their overall effectiveness. Essential players in this holistic approach are front-line AI professionals. Externalities working against the shared responsibility model require identification and critical interconnection. In order to ground these aspirations the paper will conclude with the mutual responsibility/shared fairness methodology.

The rest of the paper is organised as follows. Section 2 gives an overview of the literature around 'ethical AI', we summarise some thematic controversies in order to identify the role that ethics can play in the governance of AI and ground these themes within the context of AI development in Singapore. Section 3 offers discussion of the theoretical foundations that underpin our methodology of ethics as a shared dialogue. Section 4 delves into our proposed experimental method of 'Shared Fairness' and discusses our initial observations about the utility of the method and its current limitations. Section 5 concludes.  

\section{Ethical AI}

Organisations have responded to the increasing chorus of concerns around harms posed by AI and algorithmic systems by publishing documents outlining principles that guide their AI development and use. In recent months, much work has been done to track and compare these documents and principles. Jobin, Ienca and Vayena review 84 of such documents, and found an emerging convergence around six principles: transparency, justice, fairness, non-maleficence, responsibility and privacy; but nonetheless note that there remains substantive differences in their interpretations and methods of implementation\cite{jobin2019global}. Another study comparing 36 of these documents similarly found an emerging consensus around eight themes: privacy, accountability, safety and security, transparency and explainability, fairness and non-discrimination, human control of technology, professional responsibility, and promotion of human values\cite{fjeld2020principled}. There remains room for looking behind these league tables and exploring the reasons for priority and whether these connect with degrees of take-up and operational relevance. 

Amid the widespread approval and adoption of these principled approaches, a notable line of critique has been the disproportionate role of industry actors in their crafting and promotion\cite{hagendorff2020ethics, ochigame2019invention}. Private companies like Google, Microsoft, IBM and Tencent have taken the lead in publishing their own ethics documents and principles\cite{jobin2019global,hagendorff2020ethics}. While it is unsurprising that companies at the forefront of AI development have a hand in shaping the debates around the very technologies they are building, it would be naive to expect that they will abide by voluntary standards in the face of market pressures and growth imperatives\cite{yeung2019ai}. The murky overlap between developer and self-regulator demand an evaluation of likely contradictions in incentives. Emerging critique in recent months has been to highlight the hypocrisy of 'ethics washing', where industry players hide behind the promotion and marketing of 'Ethical AI' as a form of principled self-regulation, which then functions as an alternative to legislation and other harder-edged regulatory intervention\cite{ochigame2019invention,hagendorff2020ethics,yeung2019ai}.  

To answer such concerns, industry alliances with powerful consolidated messages are asserting a commonality of ethical imperatives to address cut-throat market risk taking. The Partnership in AI, for example, has brought together an impressive collection of organisations across industry and academia with the goal of advancing research and sharing insights across market rivalries. Alliances like these smooth over public concern by suggesting that shared ethical proscriptions will prevail in self-interested, competitive markets. This message, of ethics over profit and collaboration over market advantage, is persuasive in a regulatory climate otherwise not excited by sharp regulatory technologies. Nonetheless, it has also been argued that such industry giants ‘highlight their membership in such associations whenever the notion of serious commitments to legal regulation and business activities need to be stifled’\cite{hagendorff2020ethics}. 

These difficult contradiction cast a shadow over the legitimacy of well-intentioned ethics codes and principles and continue to be grappled with even as the field is shifting from the 'what of AI ethics' to a more operational 'how'\cite{morley2019initial}. Here, it is clear that despite a consensus around ethical principles, we have yet to witness a convergence around ethical \textit{practices}---despite an emerging literature on technical tools for addressing common ethical challenges (see \cite{morley2019initial} for an overview of these tools). Two hypotheses may explain this slippage. The first is that the discussion of AI ethics remains too high-level and abstract, making it difficult for practitioners and technicians to see the their relevance in their daily activities\cite{mittelstadt2019principles,holstein2019improving}. The second is that there has been insufficient cross-fertilisation between ethical regulatory research in academia on the one hand, and real-life application with practitioners on the other\cite{morley2019initial}. For instance, one study found that AI developers, while aware of the ethical challenges in their work, were not organisationally supported with adequate tools or methods for addressing them in their work\cite{vakkuri2019ethically}. 

This gap between principles and practices remains one of the key challenges for shifting ethics and principled design into operational reality\cite{jobin2019global}. But it is not simply about improving the transition of ethics into product design. More than this is the need for the ‘humans in the loop’ to agree that ethics has operational advantage and as such it is as important a project requirement as any other. This endeavour is central for the project explained to follow. Work incorporating AI developers' own perspectives in the ethics debate remains currently limited and relatively under explored\cite{holstein2019improving}. That said, a number of studies have emerged recently dedicated towards incorporating the voices of this group. Veale, Van Kleek and Binns interviewed public sector machine learning practitioners working across five countries to understand how they were putting considerations of fairness and accountability into their everyday practices\cite{veale2018fairness}. Holstein and his colleagues similarly sought to reveal the challenges that private sector machine learning practitioners faced when monitoring for bias and fairness, so as to comment on their operational needs in ethical compliance\cite{holstein2019improving}. Orr and Davis sought to understand how practitioners distributed responsibility across the design of their AI systems, thus focusing in the personal and collective perspectives of practitioners to highlight where they saw themselves (in responsibility terms) regarding other stakeholders in the AI ecosystem\cite{orr2020attributions}.

\subsection{Singapore's Approach to Ethical AI: The Model AI Governance Framework}

Similar concerns about the abstract nature of principled-based guidance and the difficulties of translating concepts into practice might be raised when one considers Singapore's approach to ethical AI. Here, the state-direction approach to ethical AI development has been led by the country's data protection agency, the Personal Data Protection Commission (PDPC). The PDPC has so far released two editions of its Model Artificial Intelligence Governance Framework (the Framework). With the release of its second edition, the agency also included a checklist-style guide\cite{isagopdpc} for private organisations to assess their governance practices and a collection of use-cases detailing how local and international organisations have aligned their own AI governance operations with those suggested by the Framework\cite{pdpcusecases}.

The Framework has been devised to be accountability based: voluntary in take-up and compliance while also intending to help "frame discussions around harnessing AI in a responsible way"\cite{modelaigovframework}. According to the Framework, decisions made by systems that use AI should be explainable, transparent, and fair and all AI solutions should be human-centric: taking into account the interests, well-being, and safety of human beings\cite{modelaigovframework}. The Framework was designed to help “translate ethical principles into pragmatic measures that businesses can adapt”\cite{modelaigovframeworkVERONE}. It does this through identifying four points of intervention in a business's AI deployment process where specific practices might be made to operationalise these principles: its overall governance structure, level of human oversight in an AI-assisted decision, operations management, and stakeholder communication. 

For now, it remains an open question as to how well its underlying rationale and guiding principles of explainability, transparency, and fairness have been understood by AI developers. Our initial conversations, outlined below, suggest moderate understanding and reserved uptake which potentially impedes the voluntary compliance approach advocated by the government. The challenge of operational relevance for project teams is exacerbated by the reality that the document remains directed at organisations as a whole, but has not filtered down to individuals within them who occupy positions carrying the lion's share of decision-making capabilities. Higher order managerial positions are intended to institute the framework’s governance strategies, adjusting them to “ensure robust oversight over an organisation’s use of AI”\cite{modelaigovframework}. While this is an understandable focus in a hierarchical implementation mode, one potential consequence, as alluded to above, is that this approach isolates its audience and has limited immediate operational application for other no less significant parties involved in the deployment chain, including engineers and developers building AI software, as well as end users.

\section{Ethics and Applied Research in AI development}

In line with the studies above, we have designed a methodology to elicit input from AI practitioners. This group is unquestionably essential to the development of AI products, and are often involved in choosing and applying ethical tools in their projects, as well as prioritising an ethical language in their project's design and security. Other recent studies have similarly argued for the inclusion of practitioners within the strategy-development process as a means to achieve responsible principled-design. Madaio and his colleagues, for example, iteratively co-designed a fairness-focused checklist together with AI practitioners – thus enabling them to understand both practitioners needs as well as the overall efficacy of such checklists within the wider organisational structures of companies\cite{madaioco}. The approach has two distinct advantages when promoting holistic ethical engagement: first, it addresses the current gap in perspectives from developers on the ground and approaches the discussion from an understanding of what they need; second, like Orr and Davis’s work\cite{orr2020attributions}, it helpfully situates the discussion of ethics and responsibility beyond that of a single individual. This line of research findings informs our empirical approach by validating the need for front-line inclusion to achieve a more holistic approach to ethical governance throughout the AI ecosystem, and it assists our argument for the operational importance of mutualised responsibility. 

Our approach to ethics as an operational and inclusive language as well as a normative regulatory frame requiring shared responsibility resonates with Habermas’s discourse ethics paradigm\cite{habermas1991structural}. Habermas argues that norms emerge from rational-critical deliberation: an inclusive process where opposing views are shared, and parties take part in a reasoned, reflexive, and coercion-free dialogue which ends with an agreeable decision. In any such ‘conversation’, openness (and, as we highlight later, moderating organisational power impediments) is essential if a safe space for mediated decision-making outcomes is to be possible. It would seem from the top-down, set in stone approach to AI ethics broadcasting there has been little internal debate and discourse negotiation. For our purposes the Habermasian model of optimum engagement is the mirror to reveal what does not seem to be happening in the AI ethics transposition, for most companies and state agencies who support principled ethics frames. There are also examples of this discourse of negotiation and meaning sharing in the preparation of best practice guidelines when codes of conduct are struck to encourage internal industry regulatory compliance\cite{braithwaite1982enforced}. There have been assertions from the major advocates of ethics regulation in AI that the development of their principles have been road-tested within the company culture\cite{johnsongoogle}. However, this appears to be more a validation process than any genuine debate about what should or should not stand as an ethical motivator. Ethics as a discourse, on the other hand, functions as a \textit{process} of communication which informs decision-making. Dialogue, especially moderated by outsiders, becomes a form of \textit{provocation} that interrupts the everyday practices of AI developers and helps to question practices which may have been otherwise taken for granted\cite{pangrazio2017exploring}. This, in turn, can act more sustainably and iteratively for identifying and scrutinising embedded assumptions and norms. This approach thus complements the existing top-down guideline approach by encouraging AI practitioners to engage with principles more meaningfully, and thereby 'own' the outcomes of their decisions.  

%Dialogue, directed by outsiders, as such, becomes a form of \textit{provocation} that disrupts practices. At present top-down ethical principles tend to replace the role of language and communication in operational decision-making. 

\section{Contextualised decision-making and Shared Fairness}

The aim for our method is to examine the way practitioners, team leaders, and project security advisors attribute and distribute (or do not) ethical responsibility for the AI applications they create. Many of the central and oft-recalled ethical standards lack focus: for example, who exactly should AI developers be accountable to? If AI gives material form to social practices and processes, for instance algorithmic decision-making processes entail a degree of commercial secrecy and mathematical bewilderment, then uncovering algorithmic authorship and ensuring technical transparency for the purposes of accountability is rarely sufficient alone, or feasible as a targeted strategy to pinpoint attributions of ethical accountability. The ‘Shared Fairness’ project is interested to assess practitioners perceived ethical responsibilities over key value-laden operational decisions, to identify when these decisions arise and what ethical challenges they confront, and to converse in a language of ethics and responsibility which enables practitioners to internalise ethical responsibility. In directing research attention to practitioners (designers and technicians) it is intended to evaluate and educate ethical potential at an essential operational level, regarding the complex and applied anatomy of AI. Ethics, if it is to be confidently relied on as an active regulating frame, should influence each important decision in the relational construction of AI systems. Pragmatically it is also necessary to understand how the webs of deflection are created so that these issues become someone else’s responsibility. At these levels and with these insights in view, the project adds to emerging literature that seeks to both understand how practitioners are approaching the issue of ‘ethical AI’ and include them into the development of operational ethical practices and principled design\cite{holstein2019improving,orr2020attributions,veale2018fairness}.

The methodology bridges responsibility imbalances that rest in decision-making power and technical knowledge, by commencing with personal, facilitated conversations designed to return the ethical discourse to those meant to give it meaning at the sharp end of the ecosystem. By attending to practitioners, the project will better understand ethics as a socio-technical practice, working out from the assumption that as a realistic force in regulation, ethics are dynamic, evolving and interdependent.

We argue that it is important to directly address different \textit{contexts of concern} wherein the folding of algorithms and AI into so many aspects of our lives require understandings as social and market \textit{systems} rather than only talking about responsibilised technologies. In offering research locations which test the regulatory relevance of ethics at different stages of the AI/human agency interface, this project is not satisfied with operational outcomes alone. A broader recognition of ethics applications to the AI/human interface across the ecosystem is possible via initiating conversations in many different decision-sites, and thereby revealing whether ethics is or is not a dynamic influence on the social context of AI, its purposes, problems and probabilities.

The research agenda grows from a grassroots exercise first addressing mundane challenges to responsible machine behaviours managed and manipulated by human agency with identified ethical obligations. Altruism is tempered by market/social needs and ethics is, consequentially, invested with operational clout by better recognising market requirements in settings for the advancement of social good. Once recognised as counter narratives to the importance of ethics these requirements can be understood and confronted.

\subsection{Holistic Ethics: A Conversation}
The initial method advanced is one of \textit{conversations}: about roles in the creation and use of AI and big data; about whether the AI ‘language’ makes sense for this operational experience, what is confusing, what seems to be a priority, whether it is just management-speak; about the challenges which arise at particular decision points; and about the creation of a support base with a tailored language for ethics and AI that resonates in project planning, project teams, evaluation exercises, and varied experiences across the whole of the ecosystem.

In this way, the method does not intend to evaluate ethical compliance or to question professional competencies. The project is not an opportunity for organisational management to refine their training agendas or reflect on their ethics compliance expectations, although these outcomes could eventuate once our work is internalised within the participating groups and entities. Rather, we developed our role to initiate, facilitate and make sustainable conversations in which front-line practitioners, team managers and security/compliance professionals can have their say, express their concerns, identify challenges, and participate in a process of problem-solving. We envision our role as a regulatory resource, offering a safe space for interrogating ethics and principled design primarily within the context of what we express as mutual responsibility for ‘shared fairness’.

The initial phase of the project consisted of focus groups and discussion workshops with ecosystem demographics of young designers in a major multi-national technology giant, private consultancy operatives and consultants to industry, and major state-sponsored AI technology and application developers\footnote{Due to confidentiality undertakings we are not in a position to identify participants, participant organisations, dates of focus group meetings or numbers involved. Suffice to say that consistent with research expectations for focus group coverage we are confident that the scope and professional demographics of participants are adequate for pilot observations. The pilot extended from November 2019 to April 2020.}.

Each conversation was structured in three stages. The first stage involved sharing among the group each participant’s experience in working with AI and big data. This is an exercise in self-reflection and with the facilitation of the moderator, an opportunity to build trust in the personal value of the conversation. It is interesting to see how candour develops as the conversation unfolds. Following on from the sharing exercise, the conversation moves on to discussing the gaps between how participants conceive of ethics as opposed to how ethics is being presented by the emerging regulatory frame of ‘Ethical AI’, or more specifically through the training and governance operations within their organisations. At this point, having personalised where AI and big data are important to their work, participants are confronted with some ethics compendia and particular principles are discussed for their meaning and interconnections, and how these should be prioritised in project contexts is debated. The second stage of the conversation is intended to expose virtue ethics to the work experience of the participants and discuss their attitudes to its applicability, relativity and relevance.

The third stage then moves on to explore whether participants feel a sense of responsibility for principled design. Out of this consideration of mutual responsibility on a project basis, we designed hypothetical scenarios to explore how ethical challenges arise and are addressed in the development of applications and software, or the use of data in progressing their role in a team project, employing some hypotheticals. These exercises act as discussion points about routes for action or for reasons for inaction when it comes to addressing challenges and associated problem-solving. These scenarios were designed to deal with issues such as bias, data integrity, robustness, and accountability/transparency, with a prevailing and unifying focus on fairness. Finally, we talk through what ‘language’ might make ethical regulation relevant at the front-line of development and use these ideas to offer support in building mutual responsibility for shared fairness.

\subsection{Reflexive Observations}
To date we have road tested this interactive format\footnote{Due to the intervention of the COVID-19 pandemic and restrictions on movement, the personalised interaction of the focus group has been modified by slightly more moderated online conversations.} and it has revealed:
\begin{itemize}
    \item Confusion about who has responsibility for ethical practice
    \item Market pressures that are personally felt to reduce the time for thinking about these issues
    \item Genuine interest in understanding the relationship between ethical standards and principled design
    \item Uncertainty about whether and to what extent where they sit in the chain of development experiences ethical challenges
    \item Importance of ‘fairness’ as a central ethical priority
    \item Inaccessibility of ethical language and the actioning of principles
    \item Need for help in identifying ethical challenges and structuring solutions
    \item Importance of a ‘language’ that makes operational and social sense
\end{itemize}

Even though some similarities regarding the pressures surrounding ethical decision-making have recurred in all our conversations to date, there are also different priorities for different organisations depending on their market positioning, financial security, and institutional complexity. Large multi-nationals can make reputational decisions on an ethical basis which may reduce their market share in the short term, whereas smaller companies, consultancies, and start-ups much more influenced by tight profit margins and tough competition, may not have such flexibility. Bigger organisations may have designated staff and training capacity to advance ethical principles as work practice, but smaller operations will have to engage with ethics in a much more sporadic and crisis-oriented fashion. More research should be done to identify and measure such organisationally-relative market pressures, which, we suggest, will be crucial in the near-future for understanding the nature and extent of differences in ethical engagement between start-ups, MNCs, and state-sponsored agencies, especially if governance remains reliant on self-regulatory practices.  

\subsection{Limitations}

From this initial pilot phase, it is clear that this conversational method has the potential to offer assistance to AI professionals in finding a third space where their operational language can be married with that of 'AI principles' swirling around the periphery of their work-life. Nonetheless, there are several caveats that need to be discussed for this method to achieve further success. These conversations occur in a context where professionals—particularly rank and file workers—have limited control over what they are developing in terms of the extent to which they adopt responsibility for ethical operationality\footnote{This sense of hierarchical dependency has been confirmed in the empirical experience to date.}. As such, power differentials which characterise most commercial organisations, such as those in which these professionals work, militates against empowering all ecosystem participants to accept responsibility when it is distributed to them (voluntarily or through some compliance frame).

Two challenges that must be overcome for this method to produce a mutualised responsibility with 'shared fairness' outcomes: firstly, there has to be \textit{buy-in} from top and middle managers. One potential problem in gaining and maintaining managerial buy-in is that managers and leaders may be reluctant to cede power even if it results in a more effective and pragmatic distribution of ethical responsibility. This, in turn, requires a more in-depth forensic of where organisations' power structures stand and how, within their hierarchies, the organisational policy addresses ethics and governance. Perhaps a diagnostic from this ancillary investigation of power impediments to ethics free flow will be the experience of the conversations feeding into training protocols for how to better the activation and sustainability of ethics through the organisation. 

Secondly, there has to be a decision made early in the process regarding management's participation in the conversations that are held. While it is useful to appreciate first-hand the experiences of their junior staff, power hierarchies may limit the scope of the discussion and serve to inhibit conversations. Controlling who is in the room remains a challenge even without management participation, even by focusing on teams and projects. These sub-hierarchies all have leaders, and followers, and power dynamic externalities. This is where, through generating trust, a conversational ethic will emerge that has indicated on many occasions the surfacing of honest reflection over towing a corporate line.

\section{Conclusion}
This paper offers a methodology aimed at identifying a third space where the operational language of AI practitioners might be married with the abstract, high-level principles that we presently see dominating the \textit{Ethical AI} discussion. Current approaches to AI ethics, while undoubtedly well-intentioned, fail to acknowledge project priorities and pressures embedded in wider economic socio-political contexts of AI promotion, and organisational and institutional cultures that shape both opportunities and barriers to practising ethics across the AI ecosystem. These cultures and contexts, nonetheless, heavily influence the extent to which high-level principles are understood and adapted by AI practitioners, and will undoubtedly have consequences for the algorithms and models being designed under the umbrella of AI4SG and the sustainability of projects dedicated to achieving the UN SDGs. The ’shared fairness’ method is a grassroots approach to operational ethics based on dialog and mutualised responsibility. This methodology is centred around conversations intended to elicit practitioners perceived ethical attribution and distribution over key value-laden operational decision sites, to identify when these  contentious  decisions arise and what ethical challenges they confront, while engaging a language of ethics and responsibility that enables practitioners and project teams to internalise ethical responsibility.

% One drawback of this method—working out of select institutional case-studies—is that it necessarily limits empirical standardisation or representative generalisation. Nonetheless, we suggest that this particularity of research populations has enabled us to better understand ethics as a form of embedded sociotechnical practice. Assuming commonality in project team interaction, and the ethical challenges they face enables informed, critical speculation about the organisational and market forces at work either to enable or disable ethical principles as potent regulatory tools.

\bibliographystyle{IEEEtran}
\bibliography{references}
\end{document}